\documentclass[12pt]{article}

\usepackage{amsmath}
\usepackage{amsfonts,amssymb,amsthm,overpic}
\makeatletter
\newcommand{\diam}{\mathop{\operator@font diam}}
\usepackage{setspace}
\usepackage{multicol}
\usepackage{multirow}
\usepackage[compact]{titlesec}

\usepackage{epsfig}

\begin{document}

\title{\Huge{\textsc{Are four dimensions enough; a note on ambient cosmology.}}}

\author{Kyriakos Papadopoulos$^1$, Nazli Kurt$^2$, Basil K. Papadopoulos$^3$\\
\small{1. Department of Mathematics, Kuwait University, PO Box 5969, Safat 13060, Kuwait}\\
\small{2. Open University, UK}\\
\small{3. Department of Civil Engineering, Democritus University of Thrace, Greece}\\
E-mail: \textrm{ kyriakos@sci.kuniv.edu.kw}
}

\date{}

\maketitle

\begin{abstract}
The group of homothetic symmetries in the conformal infinity (the $4$-dimensional ``ambient boundary'')
of a $5$-dimensional spacetime restricts the choice of topology to a topology
under which the group of homeomorphisms of a spacetime manifold is the
group of homothetic transformations. Since there are such spacetime topologies in 
the class of Zeeman-G\"{o}bel, under which
the formation of basic contradiction present in proofs of singularity theorems is
impossible, an important
question is raised:  why should one construct a $5$-dimensional metric, in order to
return back such a topology to its $4$-dimensional conformal boundary, while
such topologies, like those ones in the Zeeman-G\"{o}bel class, are already
considered as more ``natural'' topologies for a spacetime, rather than the
artificial (according to Zeeman) manifold topology?

\end{abstract}

{\bf AMSC:} 83–XX, 83F05, 85A40, 54–XX    \\
{\bf keywords:} Ambient Cosmology, Spacetime Singularities, 
Zeeman - G\"obel topologies, Path Topology.

%%%
\section{On Ambient Cosmology and Spacetime Singularities.}
%%%

In \cite{Cotsakis1} the authors create a model of
ambient cosmology, the ``Ambient Space - Ambient Boundary''
pair, inspired by previous approaches
on braneworlds and holographic ideas, with the motivation
to describe the spacetime singularities that are predicted in the
theory of general relativity, showing that under this
model the singularities disappear and the proposition of cosmic censorship
becomes valid. The authors start from a fixed metric
in the boundary and then they consider the conformal 
structure of this boundary for constructing a $5$-dimensional
metric that will return a suitable $4$-dimensional metric
to the conformal boundary. In this way, $4$-dimensional
relativistic manifold is examined as an asymptotic and
holographic limit of a $5$-dimensional structure, the ``ambient
space'' $(M \times \mathbb{R},g_+)$ which satisfies the
$5$-dimensional Einstein equations with fluid sources and
is defined, in a local manner, in an open neighbourhood
of the ``ambient boundary'' (the $4$-dimensional spacetime
$(M,g)$ that we live in).

In \cite{Cotsakis2} and \cite{Cotsakis3} the authors show
that, by its construction, the ambient metric $g_+$ defines
a homothetic symmetry on the ambient boundary $M$ and, since
in \cite{gobel} it has been proved that the Fine topology
(as well as other topologies of the class Zeeman-G\"{o}bel)
admits the property that the group of homeomorphisms of
a spacetime (under this topology) is the group of all homothetic
transformations of the spacetime (in other words, a homeomorphism
is an isometry)),
it should be that the unique topology on the ambient boundary,
with some physical meaning, should be the Fine topology. In \cite{Singularities-ambient}
it is mentioned that there is an erratum in article
\cite{Cotsakis3}, where there is a claim that the Fine topology
does not admit ``Euclidean-open balls with their Euclidean metric''  (see page 5
of \cite{Cotsakis3}) -here we should mention that open balls should
be defined by an appropriate Riemann metric, to be more 
correct, since we refer to spacetime manifolds-; the Fine topology,
as a finer topology than the manifold topology (by its construction, 
see \cite{Zeeman1} and \cite{gobel}) contains all the open sets of
the manifold topology. So, the assertion that ``all sequences will
be Zeno sequences'' is not valid and, hence, the assertion that
the Limit Curve Theorem does not hold, under the Fine topology, 
is not valid, as well.

In the next section, we show that there are actually topologies
in the class of Zeeman-G\"{o}bel, where the Limit Curve Theorem
fails to hold, but for a different reason.

%%%
\section{The Path Topology and the Convergence of Causal Curves.}
%%%

In \cite{Low_path}, the author shows that in the Path topology 
(see \cite{Hawking-Topology}) the Limit Curve Theorem fails to hold.
Since the Path topology is finer than the manifold topology, every
manifold-open set is also open in the Path topology; hence the argument
in \cite{Cotsakis2} that the convergence of causal curves depends
on the existence of Euclidean- (manifold-, more correctly) open balls is not
valid.
The Limit Curve Theorem (under the manifold topology) states that if 
$\gamma_n$ is a sequence of causal curves, $x_n$ is a point on $\gamma_n$
for each $n$, and if $x$ is a limit point of $\{x_n\}$, then there is
an endless causal curve $\gamma$, passing through $x$, which is a limit
curve of the sequence $\gamma_n$. The failure of this theorem, in this
sense under the Path topology, is very important, because it avoids
basic contradiction arguments that are present in the proofs of 
all singularity theorems. In \cite{Cotsakis3}, the authors highlight
that such contradictions appear when one assumes the existence of
a causal curve whose lenght is greater than some maximum that
starts from a spacelike Cauchy surface with negative curvature, 
downwards to the past. Thus, if a limit curve $\gamma$ cannot
be extracted as an appropriate limit of convergence of causal
curves, one cannot speak of geodesic incompleteness.

Since the Path topology has been shown to be the general relativistic
analogue of a topology that has
been suggested by Zeeman in \cite{Zeeman1} (see \cite{On-Two-Zeeman-Topologies}, \cite{Order-Light-Cone}
as well as \cite{LimitCurve}) and since according to G\"{o}bel in
topologies of this class the group of homeomorphisms of a spacetime manifold
 is the group homothetic symmmetries, then 
a topology which could justify the construction of the Ambient Boundary -
Ambient Space pair, in \cite{Cotsakis1} could certainly be the Path topology.

We have one more objection here, though, that we will express in the next section.

%%%
\section{Discussion: Are Spacetime Singularities a Topological Effect?}
%%%

In the previous two sections we first mentioned that the existence or not
of manifold-open sets is not linked to the validity of the Limit 
Curve Theorem and also that 
the group of homothetic symmetries of the Ambient Boundary does not
restrict our choice of an ``appropriately natural'' topology to the Fine
topology; a candidate topology, where one cannot talk about convergence
of causal curves, could be the Path topology. Since the Path topology, 
though, is a challenging alternative of the manifold topology, as it has been 
strongly suggested at least by Zeeman, G\"{o}bel, Hawking, King and McCarthy in the above mentioned papers, as it embodies the causal, differential
and conformal structures of spacetime, the objection on why one should need
a model of Ambient Cosmology, adding extra dimensions, at least from a topological
perspective becomes more and more stronger. It is evident that,
since the convergence of causal curves depends on the choice of
a topology for a spacetime, the singularity problem as a whole
can be placed within a topological frame exclusively: spacetime singularities 
seem to be a topological effect. It is the topology of the spacetime
which will determine the validity of singularity theorems or 
cosmic censorship and not the examination of a $4$-dimensional spacetime 
from a perspective of extra dimensions.

\end{document}